\documentclass[twocolumn,dvipsnames]{aastex631}

\shorttitle{Curvature and Dynamics of Musca}
\shortauthors{Kaminsky et al.}
\usepackage{graphicx}
\usepackage{commath}
\usepackage{xcolor}

\begin{document}

\title{On the 3D Curvature and Dynamics of the Musca filament}

\correspondingauthor{Aidan Kaminsky, Lars Bonne}
\email{kamina@rpi.edu, lbonne@usra.edu}

\author{Aidan Kaminsky}
\affiliation{SOFIA Science Center, NASA Ames Research Center, Moffett Field CA 94035, USA}

\author{Lars Bonne}
\affiliation{SOFIA Science Center, USRA, NASA Ames Research Center, Moffett Field CA 94035, USA}

\author{Doris Arzoumanian}
\affiliation{Division of Science, National Astronomical Observatory of Japan, 2-21-1 Osawa, Mitaka, Tokyo 181-8588, Japan}

\author{Simon Coud\'e}
\affiliation{SOFIA Science Center, USRA, NASA Ames Research Center, Moffett Field CA 94035, USA}


\begin{abstract}


Filaments are ubiquitous in the interstellar medium (ISM), yet their formation and evolution remains the topic of intense debate. In order to obtain a more comprehensive view of the 3D morphology and evolution of the Musca filament, we model the C$^{18}$O(2-1) emission along the filament crest with several large-scale velocity field structures. This indicates that Musca is well described by a 3D curved cylindrical filament with longitudinal mass inflow to the center of the filament unless the filament is a transient structure with a lifetime $\lesssim$~0.1 Myr. Gravitational longitudinal collapse models of filaments appear unable to explain the observed velocity field. To better understand these kinematics, we further analyze a map of the C$^{18}$O(2-1) velocity field at the location of SOFIA HAWC+ dust polarization observations that trace the magnetic field in the filament. This unveils an organized magnetic field that is oriented roughly perpendicular to the filament crest. Although the velocity field is also organized, it progressively changes its orientation by more than 90$^{o}$ when laterally crossing the filament crest and thus appears disconnected from the magnetic field in the filament. This strong lateral change of the velocity field over the filament remains unexplained and might be associated with important longitudinal motion in the filament that can be associated to the large-scale kinematics along the filament.

\end{abstract}


\section{Introduction} \label{sec:intro}
Molecular clouds pervading the cold interstellar medium (ISM) are structured with filaments. These filamentary clouds ubiquitous in the ISM became particularly evident with Herschel and Spitzer observations, and when sufficiently dense, host pre-stellar cores \citep{Andre2014,Hacar2022,Pineda2022}. While observations towards molecular clouds show that filamentary structures undergo fragmentation into dense cores that become stars, the physical processes that drive star-formation in filaments are not entirely understood. The magnetic field structure in these star-forming filaments may play an important role in their formation and evolution. Theoretical models indicate that gravitational fragmentation may be  limited by magnetic field support, although the current observational evidence is still a matter of debate. Polarization measurements done for various filaments indicate magnetic field orientations to be either parallel or perpendicular to the filament crests depending on their column densities \citep{Planck2016M,Fissel2016,Pillai2020,Stephens2022,Pattle2022}.

Filaments embedded within molecular clouds are defined as elongated structures with a large aspect ratio as seen in 2D maps. However, the 3D geometries of many filamentary structures remain ambiguous due to the limited information about their structure in the line-of-sight (LOS). In addition, molecular clouds may also feature complex structures such as converging filament hubs \citep{Schneider2012,Peretto2014,Williams2018,Kumar2020}, 
while several dust continuum filaments in the POS consist of velocity-coherent filamentary substructures, so-called fibers. \citep{Hacar2013,Hacar2018,Fernandez2014,Dhabal2018} 

The bulk motions of gas in molecular clouds have been studied in an effort to understand the evolution of filaments, as turbulence and/or magnetic fields may induce flows both along filamentary structures and perpendicular to them. Moreover, the gravitational longitudinal collapse of filaments has been explored in a number of theoretical studies \citep{Burkert2004}, both with and without the presence of magnetic field support. In \cite{Pon2012}, \cite{Clarke2015}, and \cite{Hoemann2022}, filaments were shown to undergo a longitudinal contraction where cores at the ends of filaments move towards the filaments' centers.  Observational evidence of flows along filaments connected to hubs have been revealed in prior studies \citep{Kirk2013,Peretto2014}. Additionally, simulations explored in \cite{Kirk2015} show that the presence of an organized magnetic field may play a significant role on filamentary structure and core fragmentation. Furthermore, observations have unveiled velocity gradients perpendicular to filaments \citep[e.g.][]{Beuther2015,Dhabal2018,Bonne2020b,Gong2021}, which are generally proposed to be the result of mass accretion.

To improve our understanding of filament dynamics and geometry, we focus on Musca, which can be found within the larger Musca-Chameleon complex. Musca has a simple geometry on the POS and is located at $\sim170$ pc from the Sun \citep{Zucker2021}. Typical column densities of the Musca filament lie in the range of $N_{\textrm{H}_2} \sim {4-6} \times 10^{21}\textrm{ cm}^{-2}$. Musca also has a rectilinear shape covering more than 6 pc in length, with fragmentation occurring in the northern and southern ends. Polarization measurements made towards the Musca cloud show that the local magnetic field is orientated perpendicular to the filament crest \citep{Pereyra2004,Planck2016M}. Additionally, low-column density striations are present farther away from the filament crest that are roughly parallel to the magnetic field orientation \citep{Cox2016}. It has been proposed that Musca is relatively young, among others, because of its lack of active star formation, its quiescent/transonic dynamics and its relatively low linear mass ($\sim$ 20 M$_{\odot}$ pc$^{-1}$) which is close to the hydrostatic equilibrium value. These properties and its relative isolation in the POS make it an ideal cloud to study the formation and early evolution of star-forming filaments. 

The Musca filament has multiple pre-stellar core candidates, and is proposed to be undergoing gravitational fragmentation in the northern and southern ends of the filament \citep{Juvela2010,Kainulainen2016,Machaieie2017}. Furthermore, a smooth velocity field along the entire crest was shown in \cite{Hacar2016}, strongly suggesting Musca is a velocity-coherent structure along the LOS. Velocity gradients perpendicular to the filament crest of Musca also indicate that accretion from the surrounding striations and ambient molecular cloud is taking place parallel to the magnetic field (i.e. the path of least resistance), with the gas slowing down as it approaches the filament crest \citep{Bonne2020a,Bonne2020b}. The 3D geometry of Musca has been called into question, as an analysis of the Musca cloud done in \cite{Tritsis2018,Tritsis2022} puts forth that Musca might not be a cylindrical filament, but rather a thin sheet seen edge-on, with a length along the line-of-sight (LOS) of $\sim6$ pc. This view of the Musca cloud is in disagreement with findings from \cite{Bonne2020b} and \cite{Zucker2021}, which support that the Musca filament more or less has a similar width along the LOS as it does in the POS. The latter study shows that the Musca cloud has a curvature along the LOS, which was recovered using {Gaia} reddening data (Sec. 5.6 of \cite{Zucker2021}). Results from other research groups also conclude on the presence of a 3D filamentary structure in the Musca cloud. The significant decrease of the (dust) temperature from the ambient molecular cloud (at $\gtrsim$ 16 K) towards the central axis of the crest points to the presence of a cylindrical structure that contains cold ($\sim$ 10 K) and dense ($\sim$ 10$^{4}$ cm$^{-3}$) gas \citep{Cox2016,Machaieie2017}. The spatial separation of the cores point to gravitational fragmentation in a filament as a sheet has different fragmentation properties \citep{Kainulainen2016} and the excitation of observed \mbox{higher-J} CO lines requires the presence of dense gas expected for a 3D filament \citep{Bonne2020a}. { Lastly, there is also the presence of power laws in the N-PDF of the Musca cloud that suggests the importance of gravitational collapse in the region \citep{Schneider2022}.} Even though the ambient gas in the Musca cloud might have a sheet morphology, a variety of studies indicate that there should be a cold and dense filamentary structure embedded in the Musca cloud. This will also be addressed in forthcoming work. Therefore, we will here work under the assumption that Musca contains a 3D filament.
Understanding the 3D morphology and dynamics of the Musca cloud is an important step in uncovering its formation and early evolution. In this paper, the magnetic field orientation and gas kinematics of the Musca filament will be explored in an attempt to recover its 3D structure. Based on these results we will then further explore observations of the Musca filament at higher resolution.

\begin{figure}
    \centering
    \includegraphics[width=0.8\columnwidth]{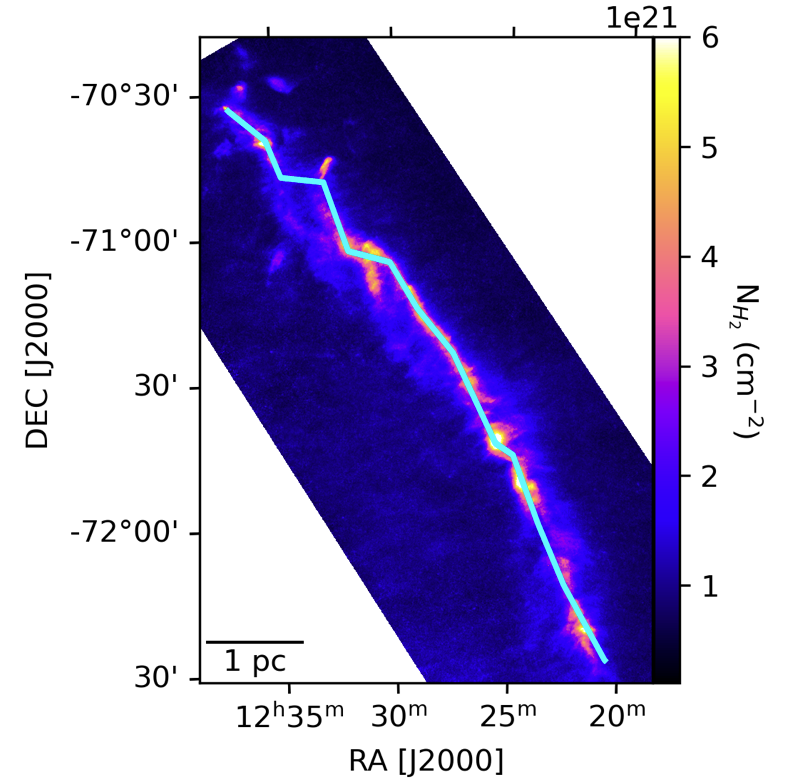}
    \includegraphics[width=\columnwidth]{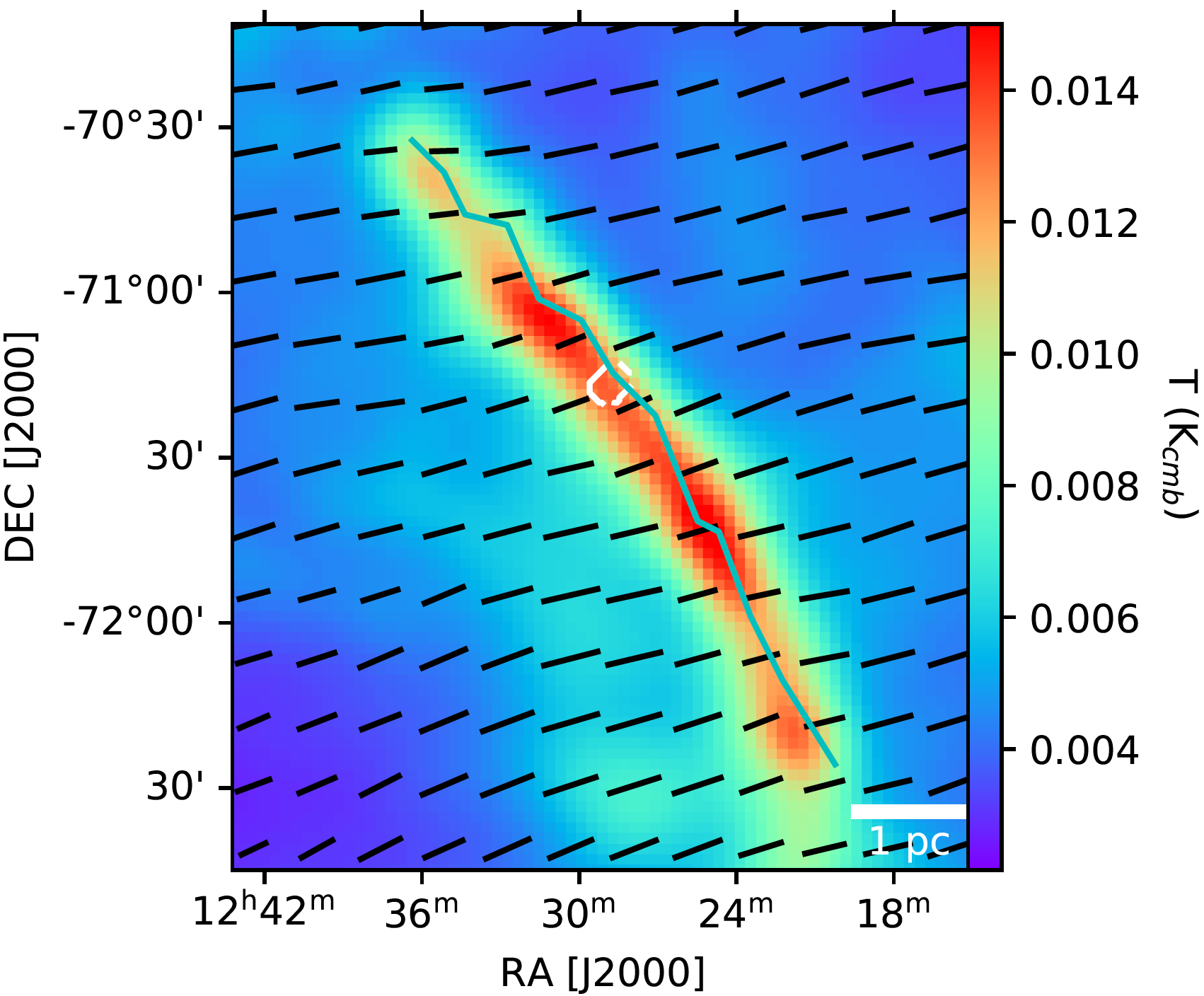}
    \caption{\textit{Top}: The filament crest of Musca as defined in \cite{Hacar2016} using dust extinction maps of the Musca cloud. The targeted positions with APEX along the crest are plotted in cyan on the Herschel Column Density map. The angular resolution of the column density map is 18.2$^{\prime\prime}$ and the spatial resolution is $\sim$ 0.02 pc. \textit{Bottom}: Magnetic field orientation derived from polarized emission measured by Planck. Magnetic field pseudo-vectors are plotted on the $I$ Stokes parameter Map, also measured by Planck. The white contour on the filament crest is the region defined by the HAWC$+$ observations. The APEX positions along the filament crest are indicated in cyan.}
    \label{fig:crest}
\end{figure}

\section{Observations}\label{sec:obs}

\subsection{APEX Observations}\label{subsec:apexobs}
We use molecular line observations along the full filament crest of Musca, specifically the C\textsuperscript{18}O($2–1$) transition (initially studied by \cite{Hacar2016}), to follow the kinematics along the filament crest. The angular resolution of the observations are 28.5$^{\prime \prime}$. These observations were carried out with the APEX 12m telescope \citep{Guesten2006}, and the data reduction and observing method is described in \citet{Hacar2016}. The crest was defined by 300 points along the filament that were at the highest column density according to dust extinction maps of the Musca cloud \citep{Kainulainen2016}, which can be seen in Fig. \ref{fig:crest}. 

\subsection{Planck Observations}\label{subsec:planckobs}
We use polarization measurements of the Musca cloud taken from the Planck legacy archive \cite{Planck2016Legacy}. The Planck $353$ GHz maps with a resolution of $4.8^\prime$ were smoothed to $10^\prime$ in order to improve the signal-to-noise ratio (S/N) of the individual \textit{I}, \textit{Q}, and \textit{U} Stokes parameter maps \citep{Planck2016M}. {The Stokes parameter maps, which are initially projected in the \texttt{HealPix} convention, and use the Galactic North Pole as a reference, were reprojected onto the equatorial reference frame. In order to calculate the polarization angles $\psi$ in this new frame, an angular correction $\gamma$ was added}. The magnetic field orientation on the POS, given by $\chi$, was computed by adding $90^{\circ}$ to the polarization angle. The equations used to determine the magnetic field angle are:
\begin{equation}
    \psi = \frac{1}{2}\textrm{arctan}(-U,Q) + \gamma
\end{equation}
\begin{equation}
    \chi = \psi+90^{\circ}
\end{equation}
The magnetic field morphology surrounding the Musca cloud can be found in Fig. \ref{fig:crest}.

\subsection{SOFIA HAWC+ Observations}
The SOFIA HAWC+ band E observations at 214 $\mu$m were centered on a single region of the Musca filament crest at $\alpha_{\textrm{J}2000}$ = 12:29:49.5 and $\delta_{\textrm{J}2000}$ = -71:16:43.7. This is located towards the central region of the Musca filament without indications of fragmentation. The observations were carried out by observing project 06\_0152 (PI: J. D. Soler), see Soler et al. (in prep.), { for a total observing time of 2h} and have a native spatial resolution of  18.2$^{\prime\prime}$ for HAWC+ band E. { The observations were carried out in the chop \& nod (NMC) mode with a nod time of of 40 seconds, a chop frequency of 10 Hz and a chop throw of 225$^{\prime\prime}$ at a chop angle of 65$^{o}$. As the Musca filament is part of the more extended Musca cloud, one has to consider potential contamination in the reference beam. Investigating potential contamination in the Herschel 160 $\mu$m and 250 $\mu$m maps \citep{Cox2016}, we find that the flux in the reference beam should be smaller than 10\%. As the Planck data shows that the variation in polarization fraction is relatively small over the Musca cloud \citep{Planck2016M}, we do not expect that the contamination in the reference beam has a major effect on the here presented magnetic field observations. Note that ideally one would investigate the Q and U maps directly for the level of contamination. However, the Planck data is at a different wavelength than the HAWC$+$ data and directly comparing the Q and U intensities would thus be based on implicit assumptions.} During the observations, 27$^{\prime\prime}$ (i.e. 1.5 times the beamsize) dithering was done to obtain data at defect pixel locations. The data we present in this paper is level 4 data from the SOFIA archive\footnote{https://irsa.ipac.caltech.edu/applications/sofia/}. The level 4 data from the SOFIA archive basically has no detected polarization at a S/N $\geq$ 3 for the polarization fraction. Therefore the data was resampled to increase the S/N. Here we have to be careful since the fully reduced HAWC+ level 4 data from the archive is oversampled with a pixel size of  4.55$^{\prime\prime}$. As a result the noise in these pixels is correlated and simply resampling the data, assuming they are independent measurements, would lead to an underestimation of the noise rms. To account for this effect, we first resampled the data to the band E detector pixel size which is 9.4$^{\prime\prime}$. For the detector pixel size it is reasonable to consider that the pixels are independent measurements leading to a $\frac{1}{\sqrt{\textrm{N}}}$ reduction of the noise level. The data was thus resampled to a resolution of 28$^{\prime\prime}$ at which point polarized emission is detected with a S/N $\geq$ 3 for the polarization fraction. It should be mentioned that the quality assessment during the observation of the project indicates that there is a small WCS uncertainty of 7$^{\prime\prime}$ which requires care when comparing the HAWC+ with other datasets. However, since we work at a spatial resolution of $\sim$30$^{\prime\prime}$ the effect of this WCS uncertainty should be minor.

\subsection{ALMA Total Power Observations}
We work with C$^{18}$O(2-1) ALMA TP observations of the Musca filament from the project with ID 2018.1.01885.S (PI: D. Arzoumanian) that covers the HAWC+ map. These observations have a spatial resolution of $\sim$30$^{\prime\prime}$ and a spectral resolution of $\sim$0.04~km~$^{-1}$. The observations were reduced using the standard ALMA single dish pipeline and the data quality was verified. We only present the single dish (or Total Power) data here as these observations have a very similar resolution compared to the HAWC+ data. For a more detailed analysis of the full data set we refer to the forthcoming papers.

\section{Large-Scale View of the Musca filament}\label{sec:res}

\subsection{C\textsuperscript{18}O Velocity Structure Along the Filament Crest}\label{subsec:c18ohacar}
Spectral line observations from the APEX 12m telescope were invoked in order to determine the LOS velocity at each point along the filament crest of Musca. In order to reconstruct the velocity field, the C\textsuperscript{18}O(2-1) spectra along the filament were fitted with a single-peaked Gaussian function, with peak intensities below $3\sigma$ ($\sigma$ = the noise rms) being removed from the dataset. 

\begin{figure*}[ht!]
    \centering
    \includegraphics[scale=1.1]{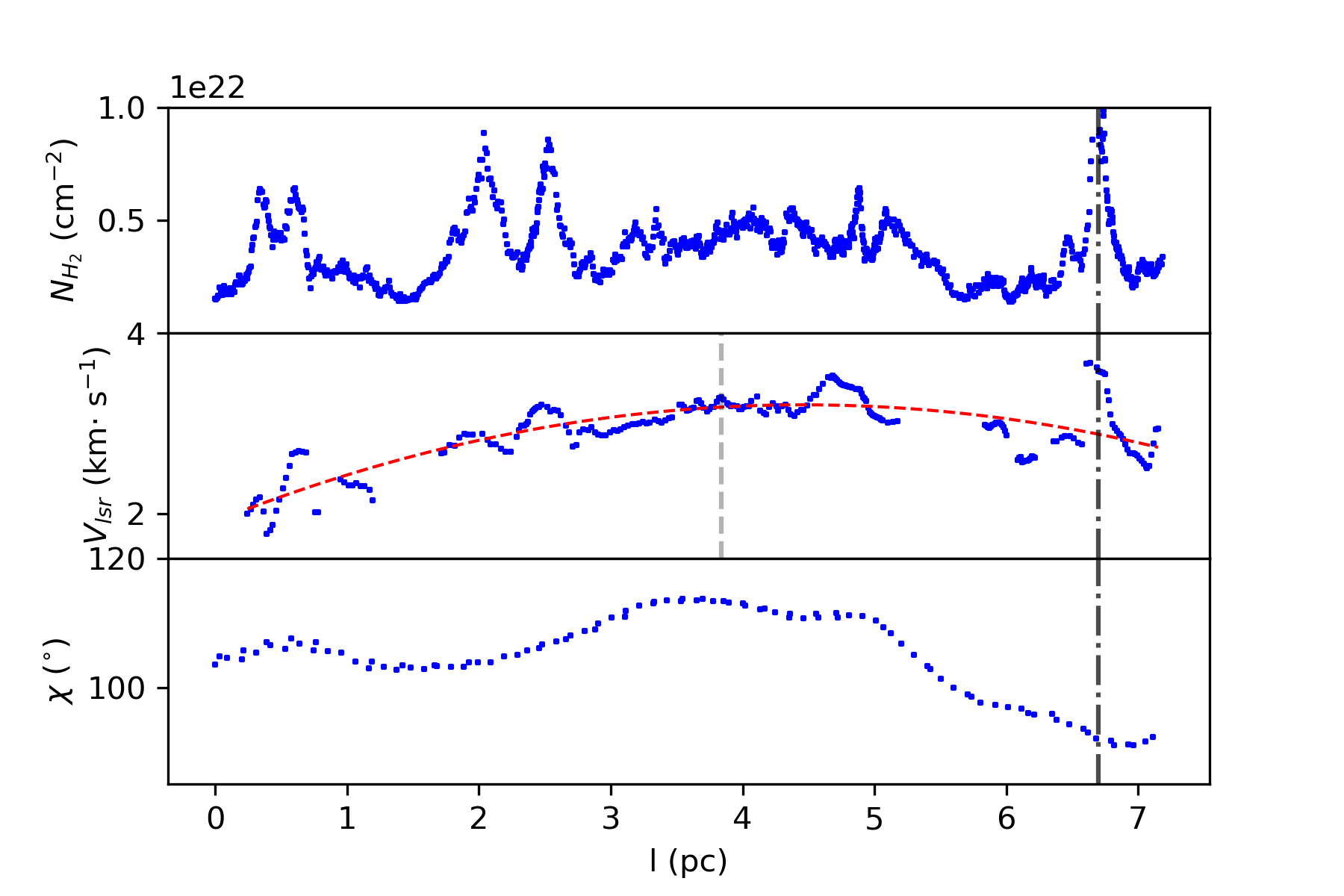}
    \caption{Longitudinal profiles of the Herschel column density (\textit{top}), the C\textsuperscript{18}O(2-1) centroid {(average)} LOS velocity ($V_{\textrm{lsr}}$) (\textit{middle}), and magnetic field orientation angles from Planck (\textit{bottom}). The angular resolutions for the column density, LOS velocity, and magnetic field angle profiles are 18.2$^{\prime\prime}$, 28.5$^{\prime\prime}$, and 10$^{\prime}$ respectively. The velocity profile was made with an S/N cutoff of 3. The dashed red line in the velocity profile is a fitted 2$^{\textrm{nd}}$ order polynomial, and {vertical dashed grey line indicates the center of the filament ($l_{\textrm{cent}}$) used in Sec. \ref{subsec:infvelmodels}.} Additionally, the position of the YSO located at the northern end of Musca is shown with the dark dash-dotted line. The filament crest used to define the longitudinal profiles can be found in \cite{Hacar2016} and Fig. \ref{fig:crest}.}
    \label{fig:int-vel-b}
\end{figure*}

The C\textsuperscript{18}O spectral data along the filament reveal a velocity-coherent structure, with a smooth velocity field over the entire crest of Musca (Fig. \ref{fig:int-vel-b}), as in \cite{Hacar2016}. Interestingly, in both the southern and northern regions of the filament crest are undetected regions where no LOS velocities were calculated {, located at $\sim 1.25 - 1.75$ pc and $\sim 5.25 - 5.75$ pc respectively}. While the southern gap is also found in Fig. 3 of  \cite{Hacar2016}, the northern one isn't because the data there is close to the 3$\sigma$ detection limit. This slight difference is then the result of the independent fitting routine that was used. However, overall the velocity field over the filament crest is fully consistent with the results of \cite{Hacar2016}. Furthermore, the crest in this work has a length of $\sim7.11$ pc, which agrees with the value in \cite{Hacar2016} after re-scaling the distance to $170$ pc, based on the results of \cite{Zucker2021}.

\subsection{$V_{lsr}$ and Magnetic Field Angle Correlation}


Using the crest as defined in \cite{Hacar2016}, see Fig. \ref{fig:crest}, the magnetic field orientation values and LOS velocity profile along the filament crest can be studied. Inspecting the results in Fig. \ref{fig:int-vel-b}, both longitudinal profiles display a smooth and apparently similar behavior along the filament crest. Specifically, the velocity field shows decreasing velocities further away from the center of the filament, while the magnetic field orientation angle similarly shows a decreasing position angle further away from the center of the filament.
It has to be emphasized that on a first consideration, there is no reason to expect a correlation between these two independent values. Even more so, as the magnetic field orientation angle only traces the morphology in the POS while the spectral lines are exclusively tracing the LOS velocity. However, as the magnetic field orientation is known to be almost perfectly perpendicular to the filament crest \citep{Planck2016M,Cox2016} the smooth change in orientation in fact is an indirect probe of the curvature of the filament in the POS. The fact that the LOS velocity field of the filament crest appears to be correlated to the smooth POS curvature of the Musca filament is interesting. This might imply that the continuum observed Musca filament is in fact the projection of a spatially curved 3D filament. This could fit with the 3D morphology at 1 pc resolution, deduced from {Gaia} reddening, for the Musca cloud in \citet{Zucker2021} that show indications of LOS distance variations for the densest gas in the cloud. However, to properly assess this, and understand the dynamics in the filament, the velocity field has to be studied with different models for the 3D velocity field.

\section{Analysis} \label{sec:analysis}
Using the C\textsuperscript{18}O spectral data first introduced in Sec. \ref{subsec:apexobs}, we attempt to construct a consistent scenario for the 3D geometry of the Musca filament. To do this, we will assume that the velocity profile along the filament is predominantly associated with longitudinal kinematics in the Musca filament. We make this assumption because the velocity difference is $\ge$ 1 km s$^{-1}$ along the filament crest. If these kinematics would be associated with bulk motion perpendicular to the 3D filament spine, this would lead to a dispersal timescale of the $\sim$ 0.1 pc POS width of the filament in $\le$ 10$^{5}$ yr. Since the filament has a well organized spatial structure in the POS, shows indications of gravitational fragmentation at both sides of the filament and hosts prestellar core candidates, this seemingly excludes such a short dispersal timescale. It thus seems reasonable at first order to assume that the observed LOS velocity profile along Musca is dominated by longitudinal motions. The reconstruction of the LOS geometry of the Musca filament will then be done by relating bulk motions along the filament to the the inclination angle at each point along Musca.

\subsection{Longitudinal Inflow Velocity Models}\label{subsec:infvelmodels}

\begin{figure*}
    \centering
    \includegraphics[width=\hsize]{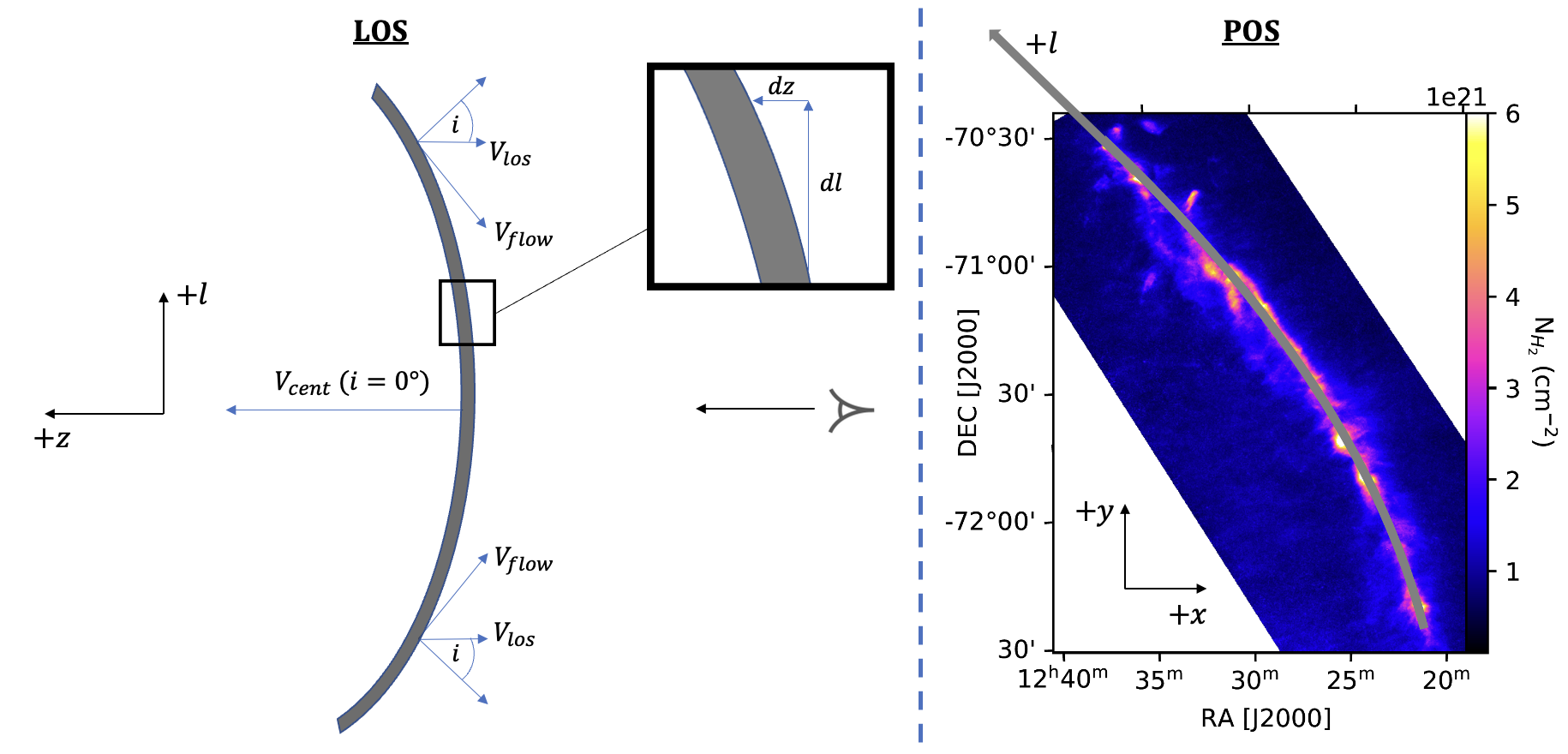}
    \caption{LOS model of Musca assuming a constant longitudinal inflow with our point-of-view indicated by the eye. The inclination angle ($i$) is measured with respect to the LOS (z-axis) and the l-axis is defined by the filament crest in the POS. This model was constructed with the velocity field expressed in Eq. \ref{eq:z_inc} and is referred to by Model 1 in Sec. \ref{sec:disc}.}
    \label{fig:initvelmodel}
\end{figure*}

To reconstruct the APEX C\textsuperscript{18}O(2-1) velocity structure of Musca, multiple velocity models along the filament were constructed. First of all, a constant inflow velocity field along the spine is considered with several assumptions, namely that the inflow velocity ($V_\textrm{flow}$) along the filament crest is constant, that the bulk motions of gas are directed towards the filament center (where the gradient of the LOS velocity field is zero, see Fig. \ref{fig:int-vel-b} for a visualization), and that the filament is moving at a velocity $V_\textrm{cent}$. {Additionally, we generalize this model so that we can parameterize the initial inclination angle $\phi$ at the center of the Musca filament relative to the LOS.} This constant inflow velocity field is thus expressed as:
\begin{equation}
    \begin{aligned}[b]
    & V_\textrm{lsr} = \\
    & \begin{cases}
          V_\textrm{cent} - V_\textrm{flow}({\textrm{sin($\phi$)}} - {\textrm{sin(i)}}) \quad &\text{if }  \, l < l_{\textrm{cent}} \\
          V_\textrm{cent} + V_\textrm{flow}({\textrm{sin($\phi$)}} - {\textrm{sin(i)}}) \quad &\text{if }  \, l > l_{\textrm{cent}} \\
     \end{cases}
    \label{eq:z_inc}
    \end{aligned}
\end{equation}
In this expression, $l$ is the position along the filament and $l_{\textrm{cent}}$ is the position at the center of the filament, see also Fig. \ref{fig:initvelmodel}. For the inflow velocity in Eq. \ref{eq:z_inc}, we use a range of velocities starting at $1.00 \textrm{ km s}^{-1}$ and ending at $6.00\textrm{ km s}^{-1}$, with an increment of $0.50\textrm{ km s}^{-1}$ (this range of inflow velocities will also be used for the models defined by Eq. \ref{eq:univel}).

In the second type of model, we assume that longitudinal inflows occur from either north-to-south or south-to-north along the entire filament. This model is defined with the following equation:
\begin{equation}
    V_\textrm{lsr} =
    \begin{cases}
          V_\textrm{cent} + V_\textrm{flow}{\textrm{sin}(i)} \quad &\text{for }  \, S \xrightarrow{}N \\
          V_\textrm{cent} - V_\textrm{flow}{\textrm{sin}(i)} \quad &\text{for }  \, N \xrightarrow{}S \\
     \end{cases}
    \label{eq:univel}
\end{equation}
{where $ N \xrightarrow{}S$ and $N\xrightarrow{}S$ indicate the North-to-South and South-to-North flow directions respectively. In the previous equation, the inclination $\phi$ is not used to generalized the unidirectional velocity field due to the fact that the resulting LOS curvature of the Musca filament will not change. For the velocity fields outlined thus far, it is important to stress again the highly idealized assumption of a constant inflow velocity along the filament crest.}

{Lastly, we construct a model in which the inflow velocity is not constant, but rather increases linearly from the center of the filament. In this scenario, we make the assumption that the bulk motions of gas are towards the filament center, and we assume that $\phi = 0$. Doing this, we can simplify the velocity field equation but infer what changes we might expect for nonzero $\phi$ cases using Eq. \ref{eq:z_inc}. This model can be expressed as}:
{
\begin{equation}
    \begin{aligned}[b]
    & V_\textrm{lsr} = \\
    & \begin{cases}
          V_\textrm{cent} - (V_\textrm{flow}+\textrm{b}(l_{\textrm{cent}}-l)){\textrm{sin(i)}} \quad &\text{if }  \, l < l_{\textrm{cent}} \\
          V_\textrm{cent} + (V_\textrm{flow}+\textrm{b}(l-l_{\textrm{cent}})){\textrm{sin(i)}} \quad &\text{if }  \, l > l_{\textrm{cent}} \\
     \end{cases}
    \label{eq:z_linflow}
    \end{aligned}
\end{equation}}
{where $\textrm{b}$ is the slope of the inflow velocity along the filament, in units of km s$^{-1}$ pc$^{-1}$. This velocity field was invoked for different values of b, more specifically 0.02, 0.10, 0.25, 0.5, 1.25, and 2 km s$^{-1}$ pc$^{-1}$.}

Note that here we do not consider end-dominated collapse models for filaments because of several reasons. First of all, models of end-dominated collapse indicate that the resulting inflow velocity reaches a maximum of 0.4 km s$^{-1}$ \citep{Clarke2015,Hoemann2022} which is significantly lower than the observed velocity field in Musca. Furthermore, reaching this terminal velocity of 0.4 km s$^{-1}$ takes a relatively long time ($\gtrsim$ 2 Myr). Additionally, even if the predicted terminal velocity would not be a limitation, the end-dominated collapse leads to a rapid velocity field acceleration towards the ends of the filament. This does not fit with the smooth observed velocity field or would imply very strong local curvatures towards the edge of the filament in our models which is likely unphysical.

\subsection{Line-of-Sight Distance Reconstruction}
The LOS distance (z) was recovered using the following equation:
\begin{equation}
    \frac{\textrm{dz}}{\textrm{dl}} = \textrm{tan(i)}
    \label{eq:z}
\end{equation}
with i being the inclination angle from the velocity models (Sec. \ref{subsec:infvelmodels}). To further illustrate how the inclination angle i is derived, see Fig. \ref{fig:initvelmodel}. Both the constant inflow velocity field (Eq. \ref{eq:z_inc}) and the alternative fields (Eq. \ref{eq:z_inc} and \ref{eq:univel}) were coupled with Eq. \ref{eq:z} in order to calculate z.
To see which equations are invoked for our different LOS reconstructions, see Tab. \ref{tab:losmodeldesc}. Additionally, for the integration required to recover z, we use an initial value of $171.60$ pc in the LOS at $l = 0$. This was obtained by fitting the LOS distances derived in \cite{Zucker2021} and estimating the LOS position at the southern-most position of the Musca cloud as defined in \cite{Zucker2021}.

\begin{deluxetable}{ccc}
\tablenum{1}
\tablecaption{LOS Model Descriptions \label{tab:losmodeldesc}}
\tablewidth{0pt}
\tablehead{\colhead{LOS Model} & \colhead{Description} & \colhead{Equations}}
\startdata
Model 1 & {Constant Velocity Field} & \ref{eq:z_inc}, \ref{eq:z} \\
Model 2 & {Unidirectional Velocity Field} & \ref{eq:univel}, \ref{eq:z} \\
Model 3 & {Linear Velocity Field} & \ref{eq:z_linflow}, \ref{eq:z} \\
\enddata
\end{deluxetable}
\begin{figure*}
    \centering
    \includegraphics{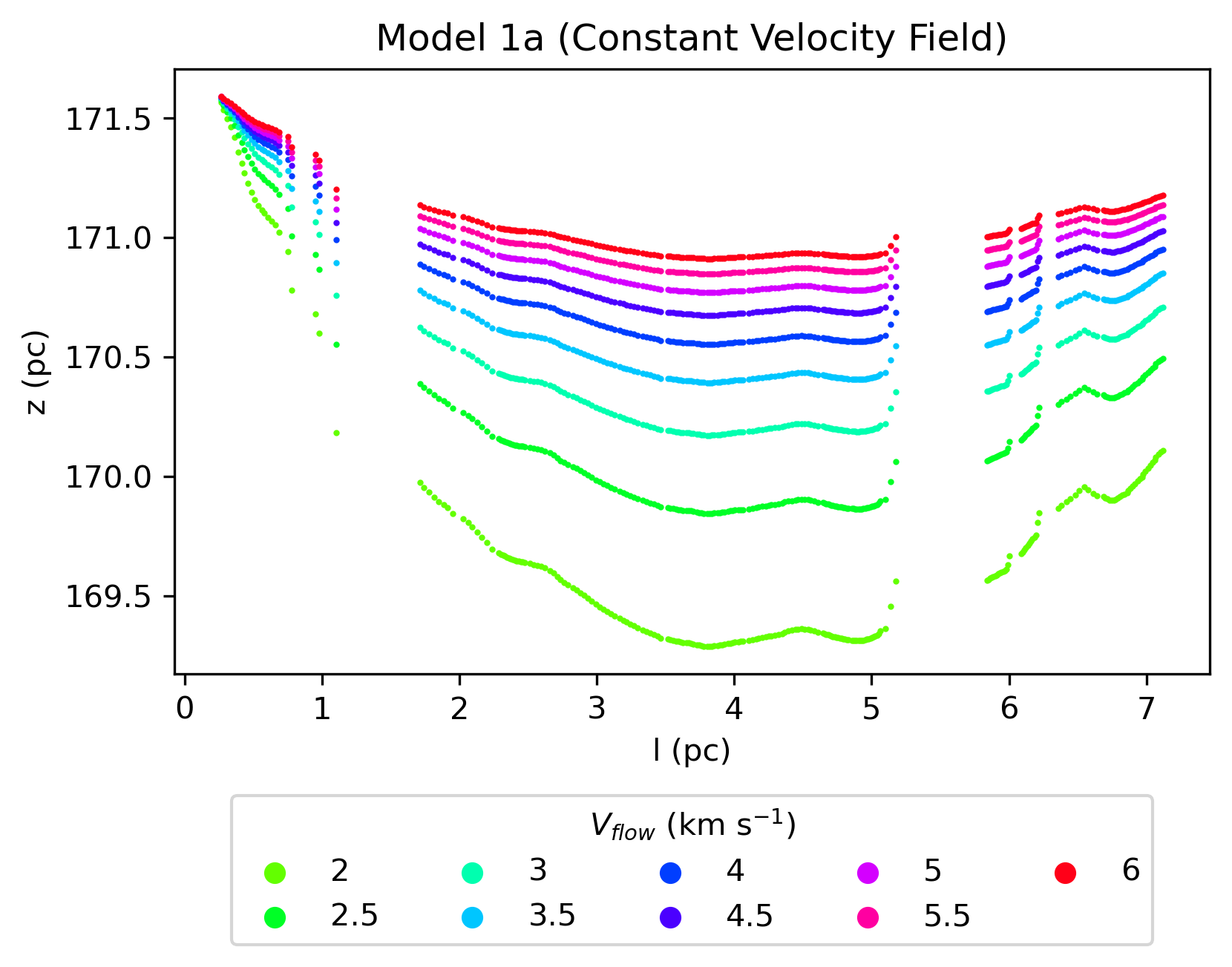}
    \caption{{LOS reconstructions using the constant inflow velocity field (Eq. \ref{eq:z_inc}) with $\phi = 0$ for different values of the inflow velocity. The curvature seen along the filament crest is similar to the geometry on the POS.}}
    \label{fig:initvelplot}
\end{figure*}
For all of the velocity fields used in our LOS reconstructions, we neglect positions along the filament crest where the absolute difference between the central velocity and the observed velocity ($|V_{\textrm{cent}} - V_{\textrm{lsr}}|$) is larger than the inflow velocity, due to the constraint that the argument inside of arcsin cannot be greater than one. As a result, we share LOS models with inflow velocities starting at $2.00 \textrm{ km s}^{-1}$ rather than $1.00 \textrm{ km s}^{-1}$. 
\begin{figure*}
    \centering
    \includegraphics[width=0.75\textwidth]{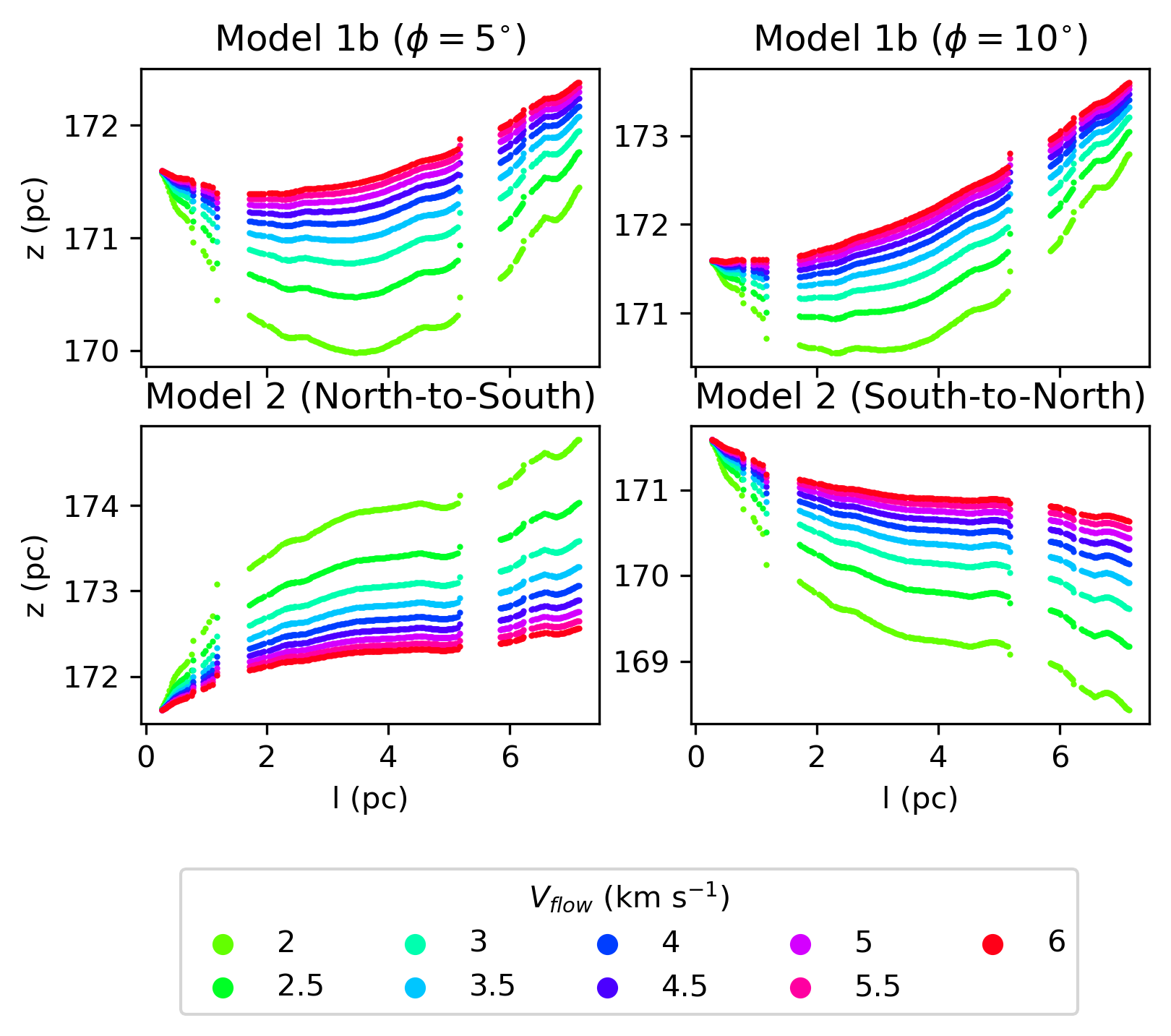}
    \caption{LOS reconstructions using the velocity field defined in Eq. \ref{eq:z_inc}, with the inclination angles $\phi = 5^{\circ}$ (\textit{top left}) and $\phi = 10^{\circ}$ (\textit{top right}). These LOS models resemble the structure of Musca as seen on the POS. Additionally, displayed are LOS models created with the velocity field in Eq. \ref{eq:univel}, which were done with a North-to-South flow (\textit{bottom left}) and a South-to-North flow (\textit{bottom right}).}
    \label{fig:incunivelplot}
\end{figure*}

\begin{figure*}
    \centering
    \includegraphics{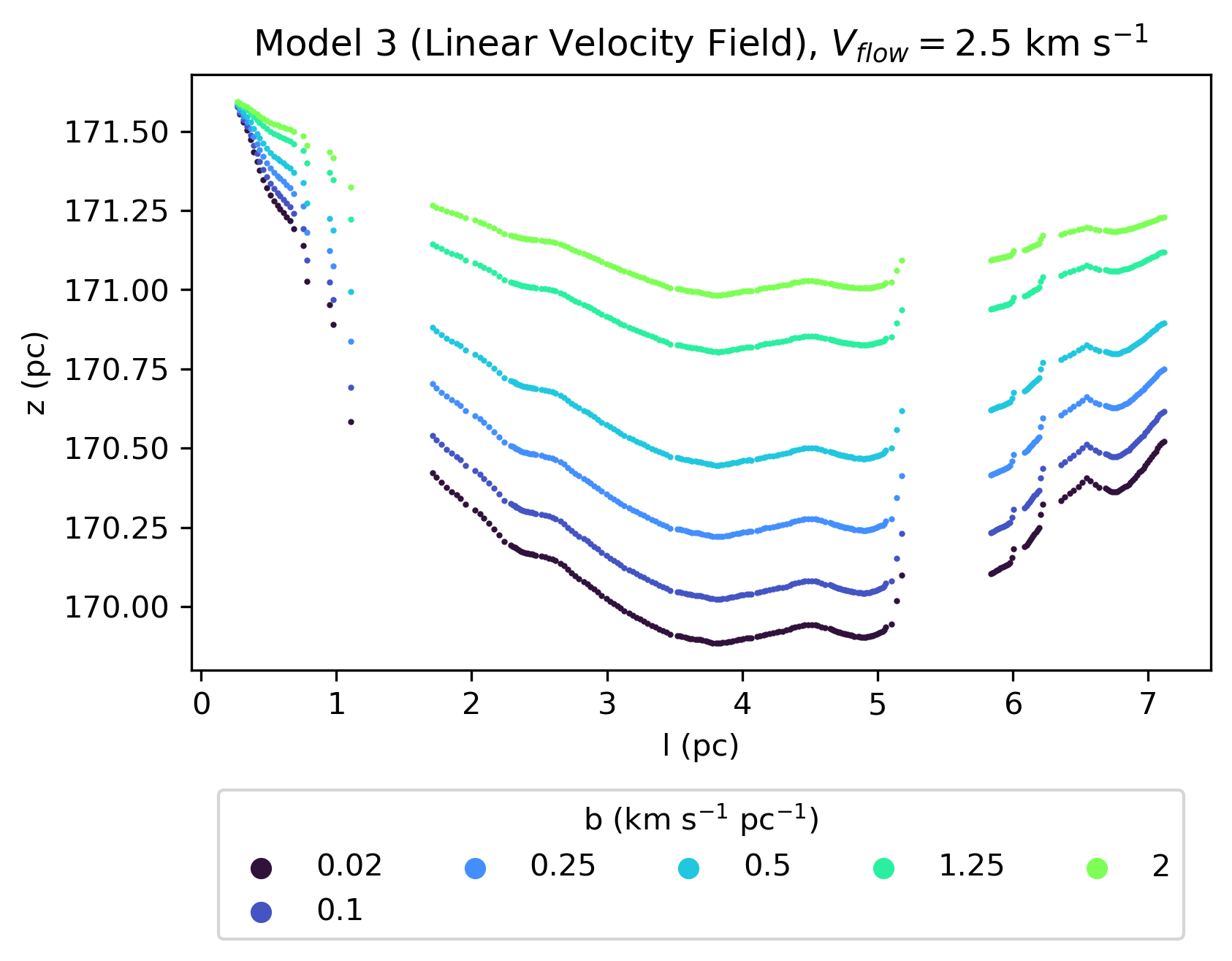}
    \caption{{LOS reconstructions using the constant inflow velocity field (Eq. \ref{eq:z_inc}) with $\phi = 0$ for different values of the inflow velocity. The curvature seen along the filament crest is similar to the geometry on the POS.}}
    \label{fig:linflow}
\end{figure*}

\section{Discussion} \label{sec:disc}
\subsection{3D Structure of Musca}

{In our analysis of the LOS models computed using Eq. \ref{eq:z_inc}, we use two variations of the constant velocity field model: one with the simple case of $\phi = 0$, and another where $\phi \neq 0$ (now referred to as Model 1a and Model 1b respectively). Both cases show a curvature along the entire filament (Fig. \ref{fig:initvelplot} and Fig. \ref{fig:incunivelplot}). While both models have a very similar curvature along the LOS, Model 1b extends further along the LOS than Model 1a, which can be intuitively explained by the slight change in orientation due to the inclination angle $\phi$. It is important to note that both Models 1a and 1b share a similar morphology to Musca as projected on the POS, which was initially suggested by the correlation found between the magnetic field orientation and the LOS velocity along the crest (Fig. \ref{fig:int-vel-b}).}

\begin{figure}
    \centering
    \includegraphics[width=\columnwidth]{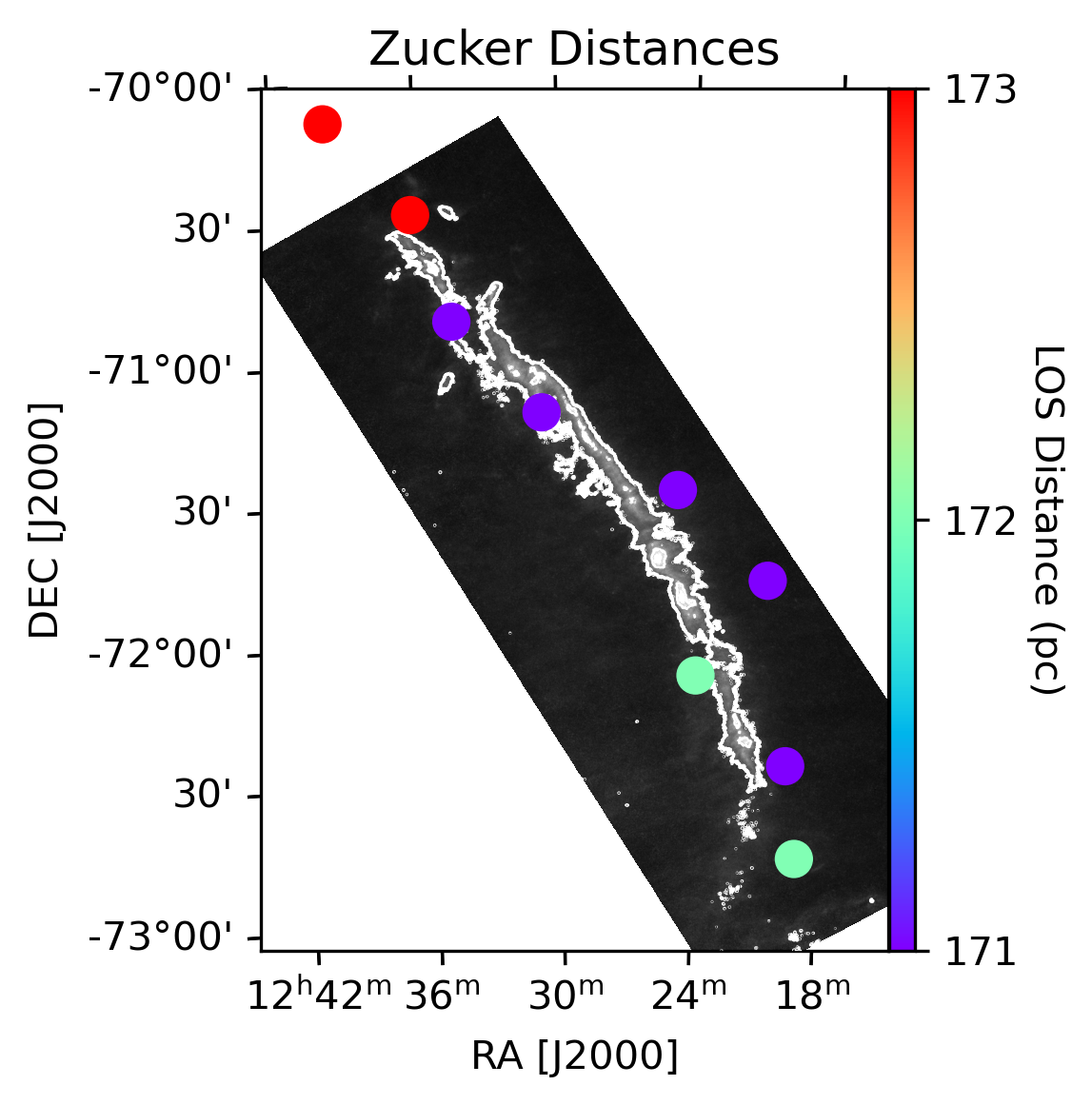}
    \caption{The Zucker Distances plotted on top of the Herschel greyscale column density map. The contour levels (in white) are made at $N_{\textrm{H}_2}$ = 2, 5, and 7 $\times$ $10^{21}$ cm$^{-2}$. The model derived in \cite{Zucker2021} can be viewed in more detail at \url{https://faun.rc.fas.harvard.edu/czucker/Paper_Figures/3D_Cloud_Topologies/gallery.html} which clearly shows the curved morphology of the region. Note that the positions are not always located on the Herschel filament crest because of the $\sim$ 1 pc resolution of the 3D extinction map compared to the $\sim$0.02 pc resolution for the Herschel data.}
    \label{fig:zuckermodel}
\end{figure}

\begin{figure}
    \centering
    \includegraphics[width=\columnwidth]{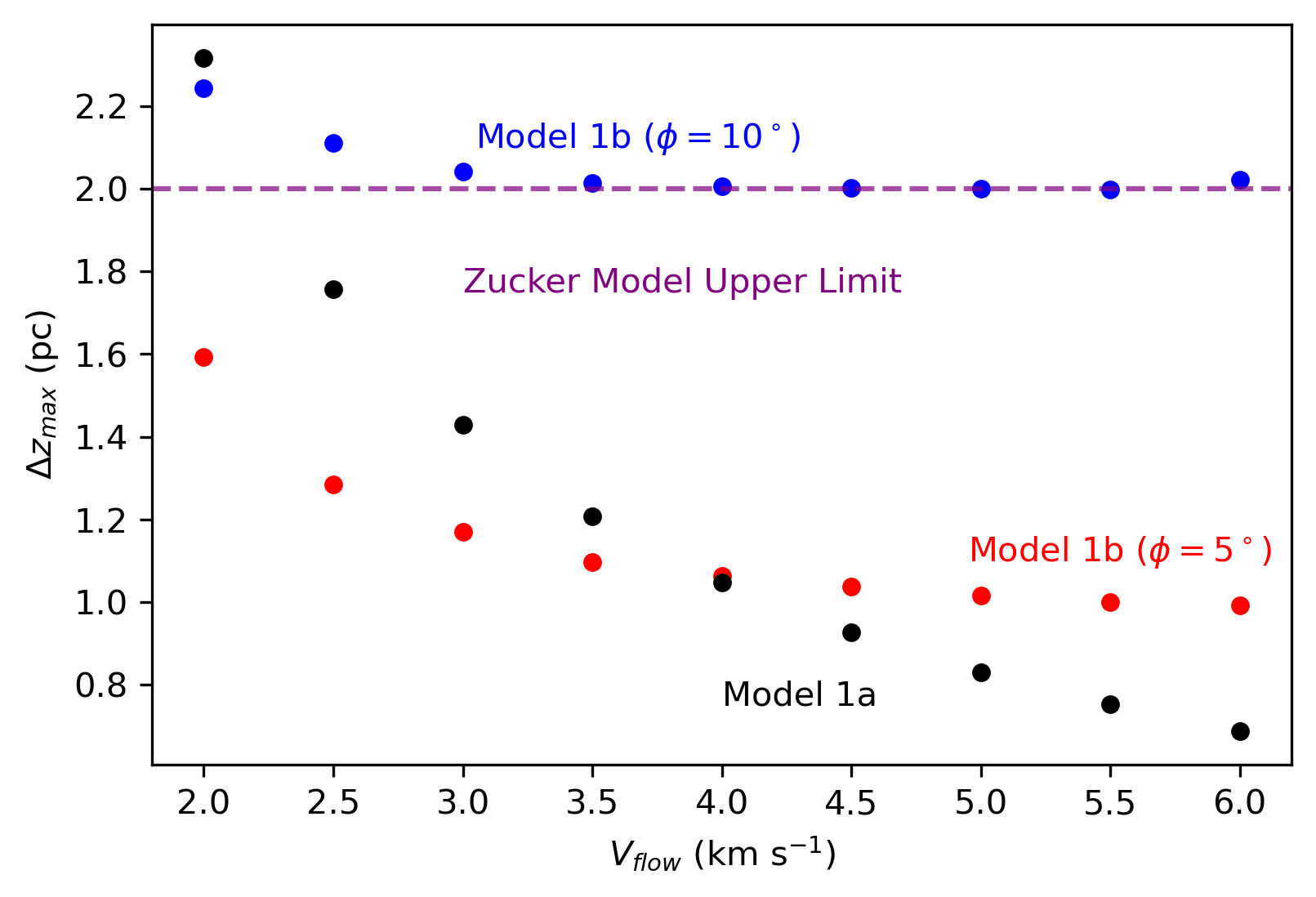}
    \caption{Maximum LOS extent ($\Delta z_{\textrm{max}}$) as a function of velocity for Model 1 (\textit{\textcolor{black}{black}}), Model 2 ($\phi = 5^{\circ}$) (\textit{{red}}), Model 2 ($\phi = 10^{\circ}$) (\textit{{blue}}), and the Zucker Distances (\textit{{purple}}).}
    \label{fig:w_max_plot}
\end{figure}

In order to further analyze our LOS models (see Tab. \ref{tab:losmodeldesc} for overview), we use the LOS distances along the Musca cloud derived in \cite{Zucker2021} (now referred to as the Zucker Distances, see Fig. \ref{fig:zuckermodel}). Because \cite{Zucker2021} uses targeted positions that do not follow the crest as defined in \cite{Hacar2016}, a direct comparison cannot be made. 
The longitudinal positions would differ in length by $\sim 1$ pc and ignoring velocity gradients perpendicular and outside the filament may not be a reasonable assumption to make \citep{Bonne2020b}. However, it is important to state that the crest we use to reconstruct the LOS structure of Musca properly traces the center of the filament and is not affected too much by these perpendicular velocity gradients. The resolution of the 3D dust map used in \cite{Zucker2021} (although initially shown in \cite{Leike2020}) is roughly $1$ pc, so the Zucker Distances do not resolve the filament considering the width of Musca is only a few tenths of a parsec \citep{Cox2016}. However, the LOS distances given in \cite{Zucker2021} do trace the peak volume density along the filamentary structure and thus the volume that hosts the filament. This justifies using the Zucker Distances, which show a clear curved morphology in their 3D online rendering, as a means of verifying the correct curvature of our own LOS models (Fig. \ref{fig:w_max_plot}).

{Models 1a and 1b} show a very similar behavior to the Zucker Distances. The Zucker Distances show a maximum LOS extent ($\Delta z_\textrm{max}$) of 2 pc, with the closest point along Musca being at $\sim$171 pc (which can be viewed as the South-Central region of Musca) and the farthest point at $\sim$173 pc (located at the northern end of Musca) (Fig. \ref{fig:zuckermodel}). This value of $\Delta z_\textrm{max}$ is an upper limit however, due to the fact that the most northern position is not covered by our velocity model and the second most northern position is associated with the protostellar core \citep{VilasBoas1994} which appears slightly disconnected from the filament in the POS (Fig. \ref{fig:w_max_plot}), but still sampled in C$^{18}$O. It is these two northern positions that give rise to $\Delta z_\textrm{max}$ = 2 pc. Models 1 and 2 show a decrease in $\Delta z_\textrm{max}$ as a function of inflow velocity, which intuitively makes sense given that the inflow velocity is inversely proportional to the inclination angle (Eq. \ref{eq:z_inc}). In {Model 1b}, the maximum LOS extent asymptotically reaches roughly $1$ pc and $2$ pc, for $\phi$ being $5^{\circ}$ and $10^{\circ}$ respectively (Fig. \ref{fig:w_max_plot}). Using the $2$ pc upper limit of the Zucker Distances, Model $2$ ($\phi = 10^{\circ}$) likely overestimates the inclination angle of Musca and thus may not reliably model the LOS structure of the filament. Using {Model 1a} gives $\Delta z_\textrm{max} = 1$ pc for $V_{\textrm{flow}}$ = 4 km s$^{-1}$ and $\Delta z_\textrm{max} = 2$ pc for $V_{\textrm{flow}}$ = 2 km s$^{-1}$ (Fig. \ref{fig:w_max_plot}). This suggests that an inflow velocity between  $2\textrm{ km s}^{-1}$ and $4\textrm{ km s}^{-1}$ would result in a LOS model most like the Zucker Distances. 
We also note that the curvature range for these models in the LOS are very similar to the ones observed in the POS. The global position angle of the Musca filament in the POS varies between 40$^{\circ}$ and 75$^{\circ}$ (or a curvature range of $\sim$ 35$^{\circ}$), while the {Models 1a \& 1b} that are consistent with the Zucker Distances have a curvature range up to 30$^{\circ}$ - 50$^{\circ}$. The curvature in the POS and the LOS are thus of the same order. The similarity of both {Models 1a \& 1b} to the Zucker Distances thus hint towards a global inflow, with a small average inclination angle in the LOS. {It is important to note that Model 1b was also constructed using $\phi$ values of $-5^{\circ}$ and $-10^{\circ}$, however orientations along the LOS with these negative $\phi$ values result in LOS models that are in disagreement with the Zucker Distances.}

The Unidirectional North-to-South and South-to-North Velocity Fields (Eq. \ref{eq:univel}) also display a curvature along the LOS, although the consequent LOS models (now referred to as {Model 2}) do not indicate that the assumptions made for these velocity fields are entirely reasonable (Fig. \ref{fig:incunivelplot}). {Model 2} displays a curvature unlike the Zucker Distances (Fig. \ref{fig:incunivelplot}), due to the assumption that bulk motions of gas flow from one end of the filament to the other. As a result, {Model 2} most likely does not recover the 3D curvature of Musca.

{The linear velocity field model (referred to from now as Model 3) displays a curvature along the LOS similar to Models 1a and 1b and the Zucker Distances. This similarity can be easily seen in Eqs. \ref{eq:z_inc} and \ref{eq:z_linflow}, as the only difference between the two models is the inflow velocity's growing term, which is dependent on the parameter b. For the case where the inflow $V_{\textrm{flow}} = 2.5 \textrm{ km s}^{-1}$, $\Delta z_\textrm{max}$ was determined to be 1.69, 1.56, 1.36, and 1.14 pc with respective b values of 0.02, 0.10, 0.25, and 0.50 km s$^{-1}$ pc$^{-1}$. Additionally, for inflow velocities of 2$\textrm{ km s}^{-1}$ and 3$\textrm{ km s}^{-1}$, we also find maximum LOS extents between 1 - 2 pc for this range of b values, except for the case where $V_{\textrm{flow}} = 2 \textrm{ km s}^{-1}$ and b = 0.02 km s$^{-1}$ pc$^{-1}$. For b $> 0.5$ km s$^{-1}$ pc$^{-1}$, $\Delta z_\textrm{max}$ decreases below 1 pc and the LOS structure flattens out, which can be seen in Fig. \ref{fig:linflow}, for $V_{\textrm{flow}} = 2.5 \textrm{ km s}^{-1}$. However, even for higher values of b, both the POS and LOS structure of Musca appear to be curved.}

{We have only considered constant and linearly-increasing velocity field models}, that appear to give rise to a smooth curvature that fits with the observed curvature in the POS. Considering a second, or even higher, order polynomial velocity field would allow the inclusion of accelerated velocity fields towards the edge or the center of the filament. This would not adapt the overall morphology but rather (slightly) adapt the local curvature angles. This accelerated term should not be too high as this would lead to smooth behaviour with a sudden very rapid change in curvature which is likely unphysical.

The favored velocity model that was developed is thus the inflow towards the center. This would drive mass accumulation at the center of the filament which is not particularly evident from the column density map \citep[e.g.][]{Cox2016}, although Fig. \ref{fig:int-vel-b} does show a slight increase in column density towards the center of the filament. The mass accumulation ($\dot{\textrm{M}}$) can be calculated using $\dot{\textrm{M}}$ = $\rho\pi\textrm{R}^{2}V_{\textrm{flow}}$. The density $\rho$ is given by $\rho$ = $\mu$m$_{\textrm{H}}$n$_{\textrm{H}_{2}}$ with $\mu$ (= 2.33) the mean molecular mass, m$_{\textrm{H}}$ (= 1.67$\times$10$^{-27}$ kg) the proton mass and n$_{\textrm{H}_{2}}$ (= 10$^{4}$ cm$^{-3}$) the volume number density estimated in \citet{Bonne2020b}. R (= 0.068 pc) is the filament radius (from \citet{Cox2016} corrected to a distance of 170 pc) and $V_{\textrm{flow}}$ the inflow velocity. Using an inflow velocity of 2.5 km s$^{-1}$ gives a mass accumulation rate toward the center of 21 M$_{\odot}$ Myr$^{-1}$. Multiplied by two, as inflow is coming from both sides, this gives 42 M$_{\odot}$ Myr$^{-1}$. This is significantly larger than the estimated sideways mass accretion rate of $\sim$15-20 M$_{\odot}$ Myr$^{-1}$ over the full filament \citep{Bonne2020a,Bonne2020b} which uniformly increases the linear mass in the filament and, if large enough, might mask this mass accumulation at the center. Mass accumulation at the center should thus be observed in significantly less time than a Myr in the presented case. However, the increasing inclination angle with the POS at the ends of the filament will lead to more mass in the LOS and might thus partially mask this mass accumulation at the center. We expect the mass along the LOS at the ends of the filament to increase up to 18$\%$ for realistic models. Additionally, if the filament has a small inclination angle at the presumed center, {corresponding to Model 1b with $\phi$ $\approx$ 5$^{\circ}$}, this would significantly reduce the required inflow velocity and thus the resulting mass accumulation at the center of the filament. {Furthermore, if we take the inflow velocity to increase linearly from the filament center (as in Model 3), the mass accumulation rate can be expressed as $\dot{\textrm{M}} \approx \rho\pi\textrm{R}^{2}(V_{\textrm{flow}}+\textrm{b}|\Delta\textrm{l}|)$. At the filament center ($|\Delta\textrm{l}| = 0$), the mass accumulation rate would remain the same as in the case where the velocity field is constant along the filament crest. However, it can be seen that with $V_{\textrm{flow}} =$ 2.5 km s$^{-1}$ the mass accumulation at positions out from the center of Musca have mass accumulation rates that are higher than the central mass accumulation rate. For b = 0.5 km s$^{-1}$ pc$^{-1}$, the mass accumulation rate at 1, 2, and 3 pc from the filament crest is $\dot{\textrm{M}} \approx$ 26, 30, and 34 M$_{\odot}$ Myr$^{-1}$ respectively. Additionally, for the case where b = 1 km s$^{-1}$ pc$^{-1}$, the mass accumulation rate at distances of 1, 2, and 3 pc from the filament crest is $\dot{\textrm{M}} \approx$ 30, 38, and 47 M$_{\odot}$ Myr$^{-1}$ respectively. Therefore, if the inflow velocity is not constant, Musca will have a non-uniform accumulation rate along the filament that can help to mask a rapid accumulation of mass near the filament center, which is predicted for a constant inflow velocity and not observed. It is important to note that we just show this as an example with the proposed density at a specific location in the Musca filament. We do not attempt to build a self-consistent model for the density evolution which is out of the scope of this paper. 
As the column density along the filament crest is relatively constant, this may be imply that the density is still fairly constant along the crest for the time being. This may be the result of accretion perpendicular to the filament crest.}

We note that we considered the case where the region of Musca, detected in C$^{18}$O(2-1), is a filament even though it has also been proposed as a sheet seen edge-on \citep{Tritsis2018,Tritsis2022}. This has not been fully addressed in this paper as the density structure of the Musca cloud will be the topic of a detailed forthcoming paper. We do however note here that it seems difficult for us to reconcile the velocity field with a sheet that would appear as such a narrow filamentary structure on the POS. Lastly, we also note that if Musca would be a curved sheet, that the (trans-)sonic linewidth \citep{Hacar2016} of the C$^{18}$O data at each position along the filament is somewhat surprising while the velocity field along the filament has a magnitude in the 1-1.5 km s$^{-1}$ range. If the filament is a sheet seen edge-on, one might expect a similar velocity range at each position in the LOS which would result in a typical C$^{18}$O linewidth of 1-1.5 km s$^{-1}$ at each position along the filament. Such a local linewidth along the filament is however observed for $^{12}$CO(1-0) \citep{Bonne2020b,Tritsis2022} which is sensitive to the outer regions of the molecular cloud, due to opacity effects, and might thus trace a sheet-like structure surrounding the dense filament.

\begin{figure}
    \centering
    \includegraphics[width=\columnwidth]{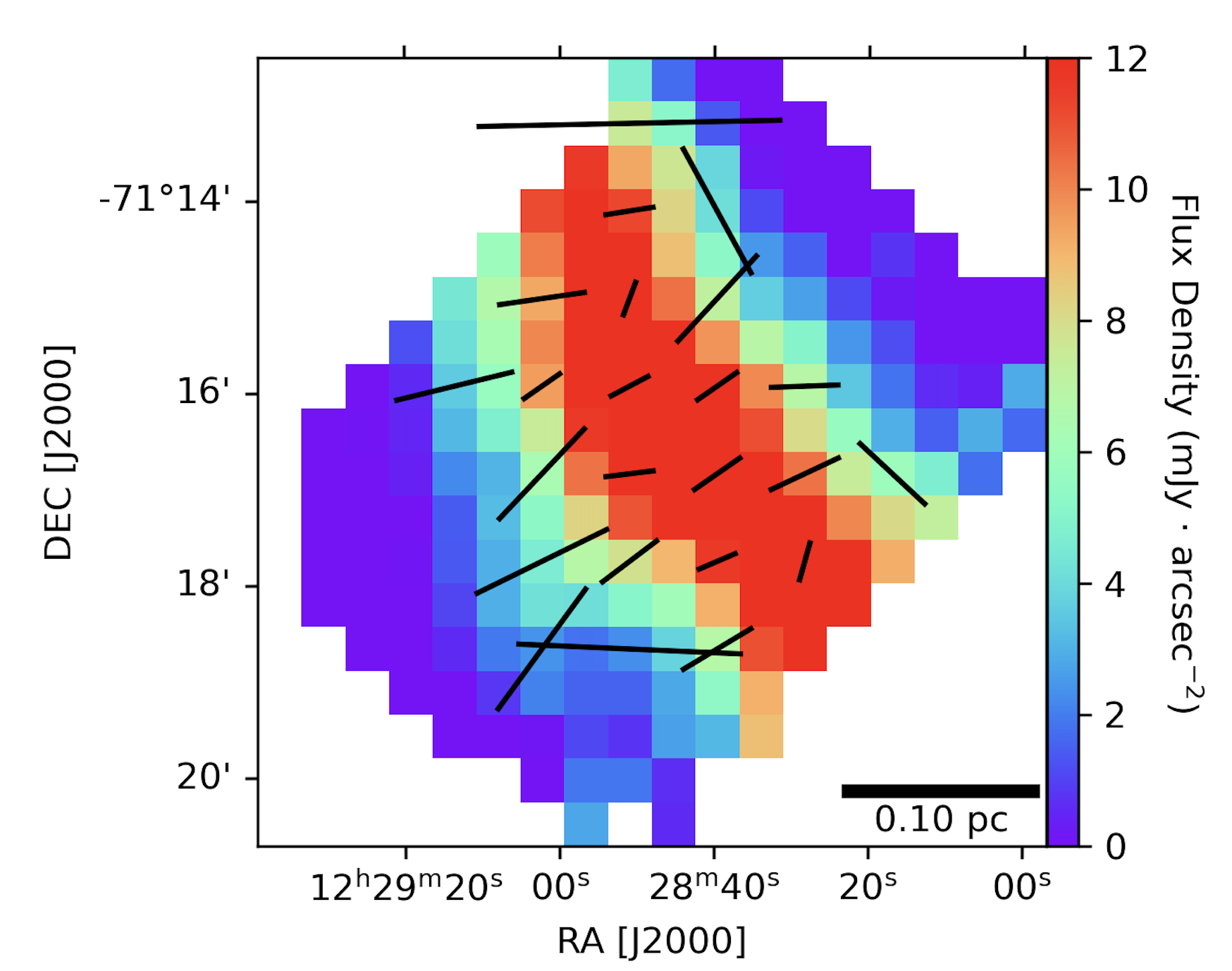}
    \caption{Magnetic field pseudo-vectors computed from the HAWC$+$ observations at 214 $\mu$m, using a S/N $\geq$ 3. The magnetic field orientation was plotted on the Intensity map of the Musca crest, which has an angular resolution of 28$^{\prime\prime}$.}
    \label{fig:SOFIA-BPOS-MAP}
\end{figure}


\begin{figure}
    \centering
    \includegraphics[width=\hsize]{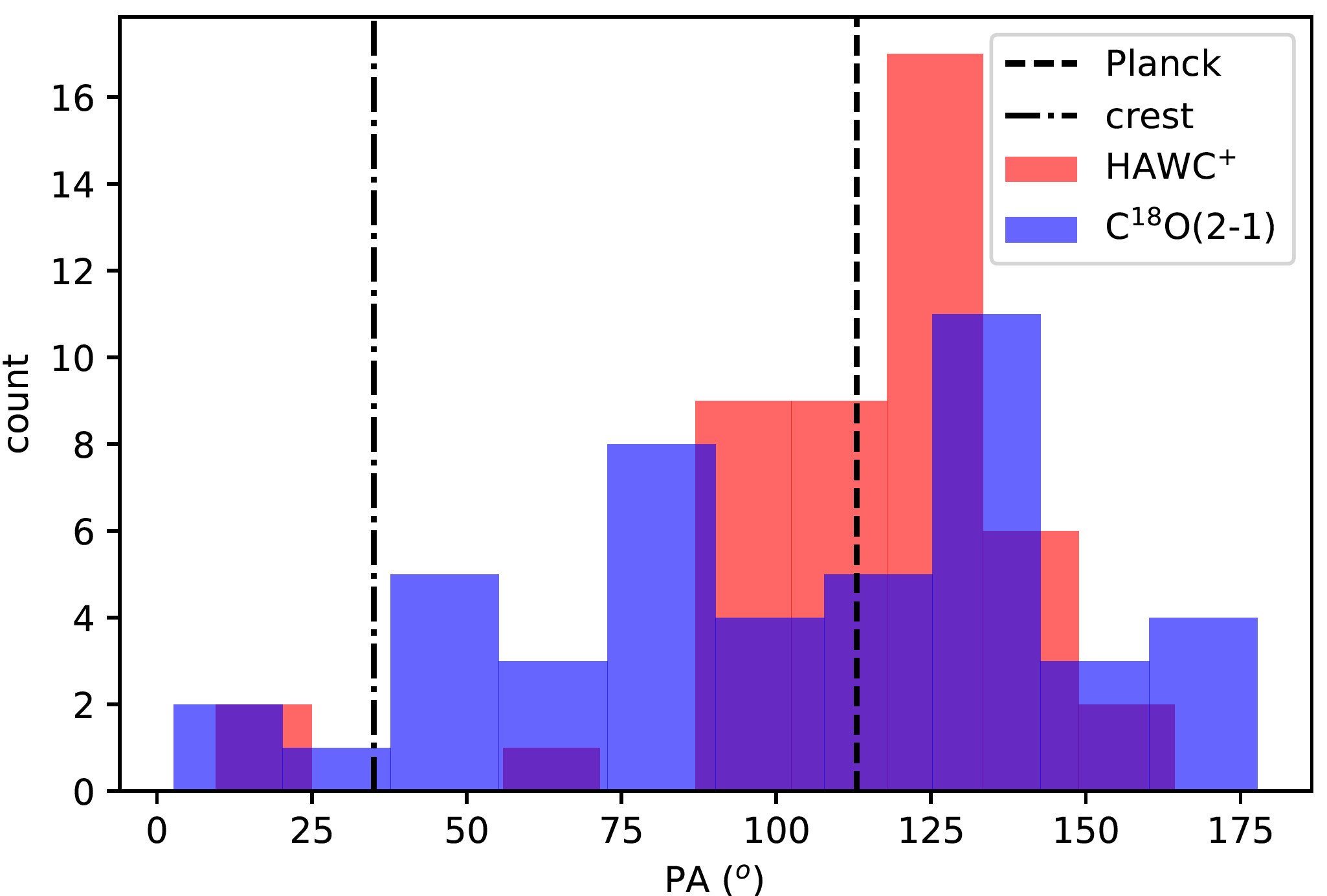}
    \includegraphics[width=\hsize]{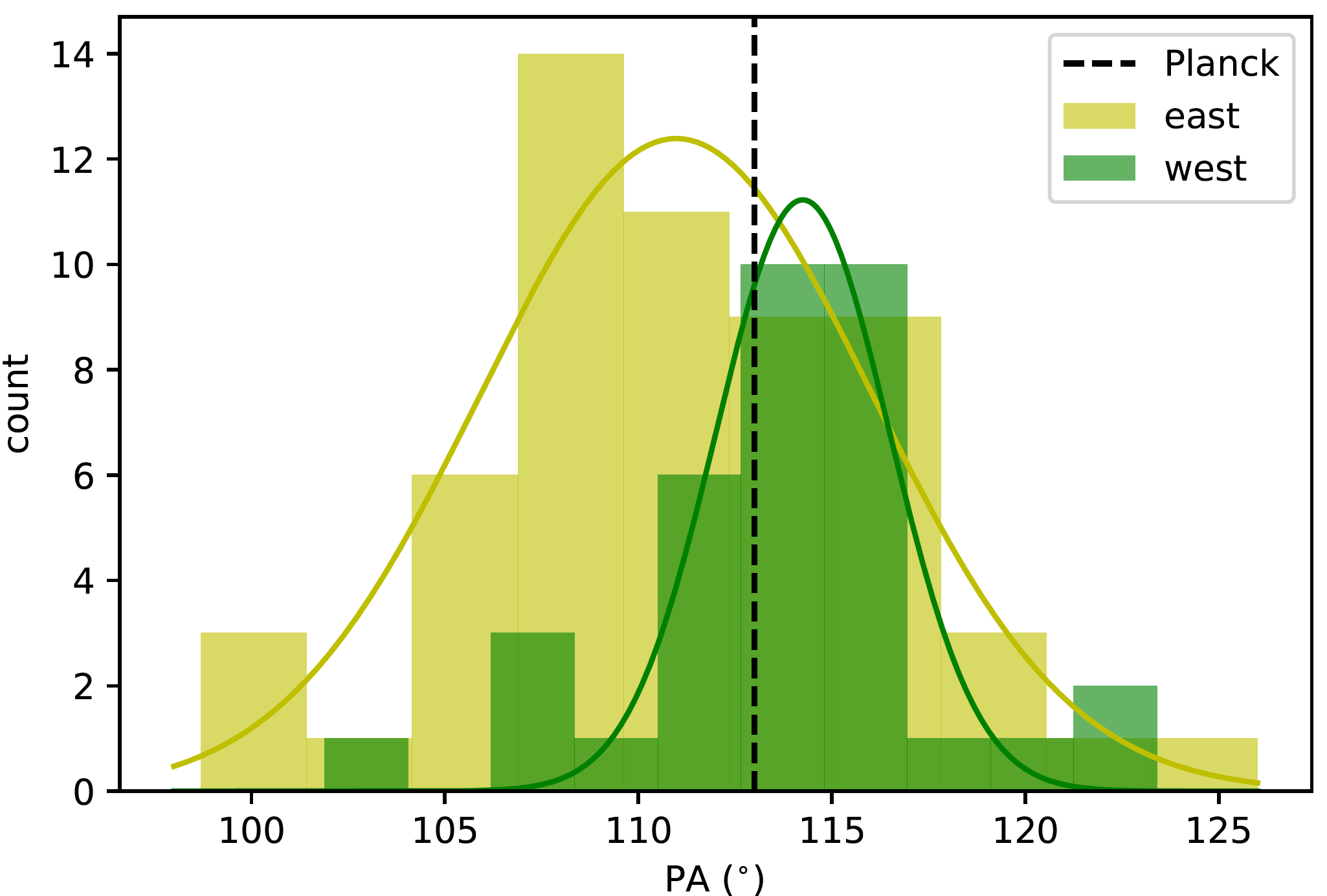}
    \caption{\textit{Top}: Histograms of the position angle (PA) for the HAWC+ magnetic field (\textit{{red}}) and the velocity gradients of the velocity field in the filament deduced from C$^{18}$O(2-1). The dashed vertical line indicates the magnetic field angle from Planck and the dash-dotted line indicates the position angle of the filament crest at the location of the HAWC+ map. \textit{Bottom}: Histograms of the magnetic field position angles from \citet{Pereyra2004} east (\textit{{yellow}}) and west (\textit{{green}}) of the filament crest. The curves indicate the result of fitting a single Gaussian to these histograms. The dashed vertical line indicates the position angle of the Planck magnetic field orientation in the region.}
    \label{fig:VELFIELD-HAWC}
\end{figure}

\begin{figure}
    \centering
    \includegraphics[width=0.8\hsize]{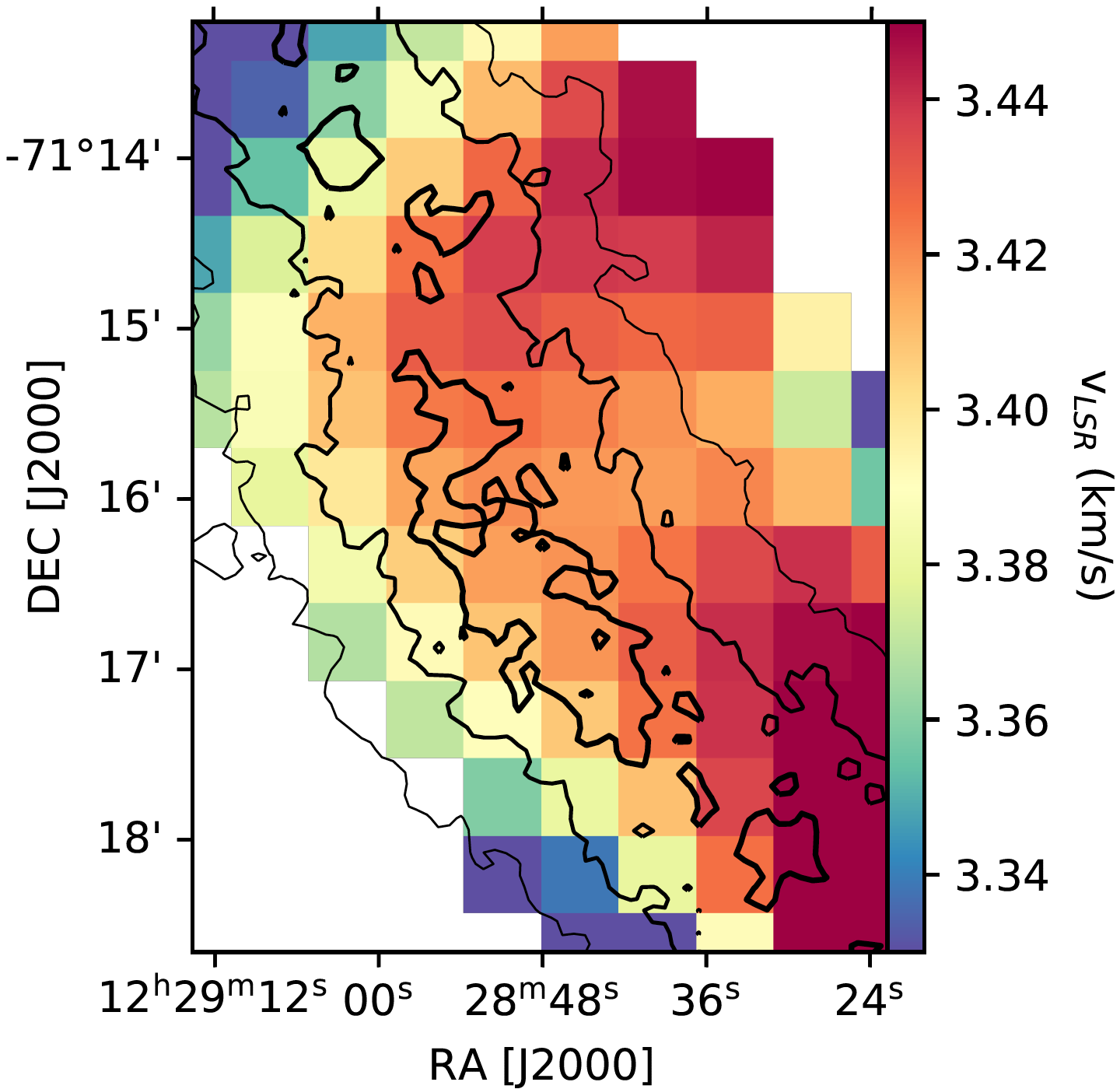}
    \includegraphics[width=0.8\hsize]{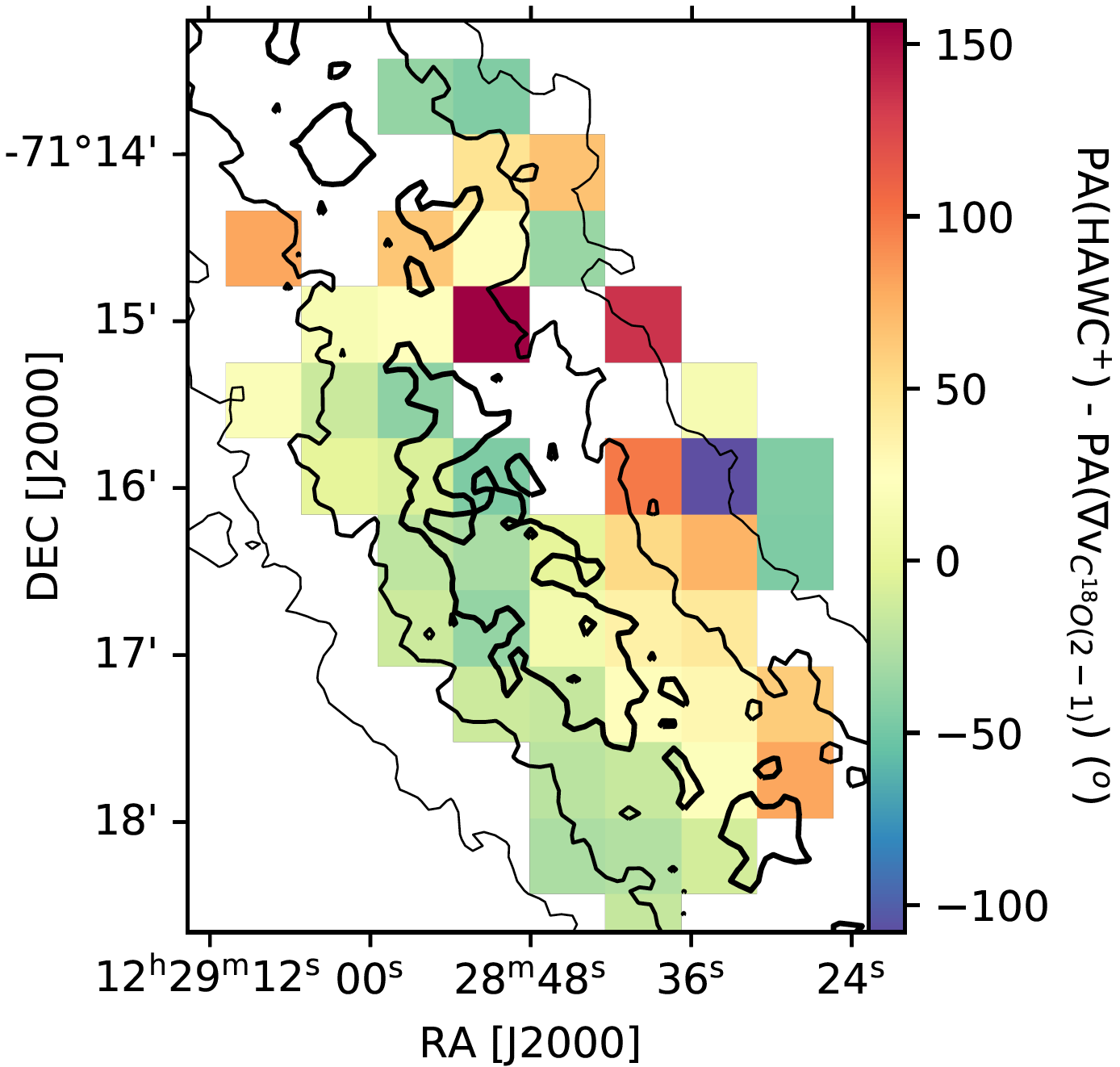}
    \caption{\textit{Top}: The velocity field at the location of the HAWC+ map as obtained from fitting a single Gaussian to the C$^{18}$O(2-1) spectra. The contours indicate the Herschel column density at $N_{\textrm{H}_{2}}$ = 3, 4, 5 $\times$ 10$^{21}$~cm$^{-2}$, highlighting the filament crest. \textit{Bottom}: The map of the local difference between the magnetic field and velocity gradient position angles.}
    \label{fig:diffVel-HAWC}
\end{figure}


\subsection{Small-Scale dynamics of the Musca filament}

In order to better understand the origin of the observed large-scale velocity field and whether it could e.g. relate to the magnetic field morphology, we here study the velocity field and magnetic field orientation at a smaller scale. As high-resolution magnetic field observations of the Musca filament crest only exist for the small region covered by the HAWC+ map, we will focus on this region. In this exact region, \citet{Bonne2020b} already noted there was a velocity gradient that was on average roughly perpendicular to the filament crest orientation. As the new ALMA Total Power data provides better coverage of the HAWC+ region, we here employ this data set.

The magnetic field orientation is presented in Fig. \ref{fig:SOFIA-BPOS-MAP} and has an average position angle of 116 $\pm$ 15$^{\circ}$. 
This results in a relative angle of 78$^{\circ}$ with respect to the filament crest. Further note that this is very close to the local Planck magnetic field position angle of 113$^{\circ}$ at the location of the HAWC+ map. A comparison with the optical polarization measurements from \citet{Pereyra2004} is a bit more involved as it only is detected outside the filament crest due to the strong extinction. Using the same declination range as for the HAWC+ map does however give a mean magnetic field position angle of 112 $\pm$ 5$^{\circ}$ for the optical data which is basically the same value as from the HAWC+ and Planck data. However, analyzing the optical polarization data separately for positions east and west of the filament crest, we do find a small change in orientation angle over the filament crest, see Fig. \ref{fig:VELFIELD-HAWC}. Fitting a single Gaussian to the east and west position angle distributions gives 111.0 $\pm$ 0.4$^{\circ}$ (east) and 114.3 $\pm$ 0.1$^{\circ}$, or a local magnetic field reorientation of 3.3 $\pm$ 0.4$^{\circ}$.

To add the local velocity field information, we fitted the ALMA data over the region of the detected HAWC+ map the same way as was done in \citet{Bonne2020b}, see Fig. \ref{fig:diffVel-HAWC}. The resulting absolute difference in the average magnetic field and average velocity gradient position angles is 30$^{\circ}$. However, inspecting the distributions of the position angles in Fig. \ref{fig:VELFIELD-HAWC} unveils an increased complexity. While the HAWC+ position angles are all localized around 116$^{\circ}$, the velocity gradient angles in the same maps show a spread over the full range of possible position angles. This indicates a more complex velocity field than observed with the magnetic field. Inspecting the map of the relative position angles for the magnetic field and velocity gradients in Fig. \ref{fig:diffVel-HAWC}, it is observed that the relative angle varies from east to west over the filament crest. In the eastern part of the filament crest, the relative position angle is typically negative while in the western part of the filament crest, the relative angle is typically positive. As the magnetic field is basically perpendicular to the filament, the larger spread in the velocity gradient position angles suggests that the local velocity field consists of both a component perpendicular to the filament crest and a component along the filament crest. First of all, it should be noted that the component perpendicular to the crest doesn't necessarily imply filament dispersal as it could simply be residual rotational motions associated with the mass accretion on the filament \citep{Bonne2020b}. The longitudinal motion, if indeed organized over the full filament, could be associated with the large-scale proposed mass inflow along the crest to the center. Although we were not able to pin down the origin of this locally observed longitudinal component, the spatial organization of the position angles for the velocity field shows that the velocity field is unevenly distributed over the filament crest and thus likely associated with shearing motion inside the filament.

\subsection{Evolution of Musca} \label{subsec:evolu}
We proposed a velocity model consistent with the 3D curvature of the Musca filament deduced from {Gaia}, where gas flows towards the center of the filament. This 3D curvature of the filament is consistent with predictions from colliding flow as proposed in \citet{Bonne2020b}, but it raises question what drives longitudinal motions inside the filament in that scenario. As discussed {(see Sec. \ref{subsec:infvelmodels})}, it is unlikely that longitudinal gravitational collapse of the filament is responsible for the observed velocity field. Here we explore potential other explanations for the observed velocity field. In a global hierarchical collapse scenario \citep{VazquezSemadeni2019}, the Musca filament is a filamentary flow that is part of a gravitational collapse initiated on larger scales outside the filament. Hierarchical collapse simulations have found larger velocity differences over filaments that can be consistent with Musca \citep[e.g.][]{Gomez2014,NaranjoRomero2022}. However, these models typically also predict significantly accelerated velocity fields towards the center of collapse for which we see no evidence in Musca. A potential explanation for this might be that we capture an early evolutionary stage in Musca.

It is of course also possible that the velocity field is not driven by gravity but rather by inertial flows. Particularly interesting in this context might be oblique accretion shocks. It was proposed for Musca that it continuously accretes mass from gas inflow that is roughly perpendicular to the filament \citep{Bonne2020b}, while \citet{Bonne2020a} also tentatively proposed the detection of oblique accretion shocks around the filament. Theoretical work by \citet{Fogerty2017} showed that oblique shocks at the edge of a filament could  strongly change the direction of the velocity field at the shock front leading to longitudinal flows along the filament, with a reorientation of the magnetic field as the inflowing gas drags the field lines. In this context it should be noted that we found a small magnetic field reorientation over the filament that was already pointed out by \citet{Planck2016M}. However, this is far from the strong reorientation predicted by the simulations from \citet{Fogerty2017}. It may be possible that the HAWC+ 214~$\mu$m and Planck observations are particularly sensitive to the warmer outer layers of the Musca filament and that polarization observations at longer wavelengths will be necessary to probe magnetic field reorientation inside the filament. It is thus not possible to conclude whether oblique shocks may drive the proposed longitudinal flow in Musca. We should also note that inflow velocities of 2 km s$^{-1}$ along the filament crest appear reasonable, but inflow velocities as high as 4 km s$^{-1}$ (required for $\Delta z_\textrm{max}$ = 1) seem unreasonably high for Musca. As indicated, a small global inclination angle for the filament might be able to resolve this. However, it should also be considered that purely longitudinal flows might not be responsible for the full observed velocity field, with the caveat that in this case the Musca filament would be a transient object.

The potential sensitivity of HAWC+ to the outer layers of the Musca filament might also explain the apparent decoupled HAWC+ magnetic field and C$^{18}$O(2-1) velocity field which might be more sensitive to the inner regions of the Musca filament. Further high-resolution magnetic field observations at longer wavelengths will thus allow to test this as well. It has to be emphasized that the difference in position angle for the magnetic field and velocity field is organized when crossing the filament (see Fig. \ref{fig:VELFIELD-HAWC}), { i.e. there appears to be a gradual change in relative angle of the magnetic field and velocity gradient when crossing the filament}. As the HAWC+ magnetic field is well organized over the filament this shows that the velocity field direction alters when crossing the filament. A way to explain this behaviour might be important longitudinal flows assumed for the velocity models. However, simulations and theoretical studies, combined with additional observations, will be necessary to truly understand the origin of this behavior.


\section{Conclusion} \label{sec:conc}
We use polarization measurements and spectral line observations made towards Musca in order to shed light on the geometry and dynamics of the Musca cloud. The magnetic field morphology permeating Musca along with the LOS velocity of the higher column density structure embedded within the cloud have led us to a 3D curved geometry for the filament. In particular, the conclusions drawn in our work can be listed as follows:

\begin{itemize}
\renewcommand\labelitemi{o}
\item C\textsuperscript{18}O(2-1) lines observed by the APEX 12m telescope first studied in \cite{Hacar2016} and \cite{Kainulainen2016} indicate a coherent structure along the LOS with a smooth velocity field. A correlation between the LOS velocity and magnetic field orientation along the filament crest was revealed, suggesting a scenario in which Musca is a 3D curved filament (Fig. \ref{fig:int-vel-b}) that is potentially embedded in a lower density sheet that is undetected in C$^{18}$O(2-1).
\item Several LOS models of Musca were created to reconstruct the APEX C\textsuperscript{18}O velocity field (Fig. \ref{fig:initvelplot} - \ref{fig:incunivelplot}), which shows a curvature along the LOS similar to the distances of Musca estimated using {Gaia} put forth in \cite{Zucker2021} (Fig. \ref{fig:w_max_plot}). This was done by assuming that longitudinal motions dominate over perpendicular motions.
\item The models that provide a plausible 3D geometry for Musca require relatively high inflow velocities ($V_{\textrm{flow}} \ge$ 2 km s$^{-1}$) and are inconsistent with semi-analytical models of isolated filament collapse \citep[e.g.][]{Pon2012,Clarke2015,Hoemann2022}. We propose this might be due to flows originating from outside the filament{, a small global inclination angle for the filament, or a linearly-increasing velocity field about the filament center} unless the coherent POS morphology of Musca has a lifetime $\le$ 0.1 Myr.
\item ALMA observations of C\textsuperscript{18}O(2-1) towards the central region of Musca reveal a local organized velocity gradient with a range of 0.1 to 0.2 km s$^{-1}$, which was also found in \cite{Bonne2020b}. Due to the similar scales of the ALMA C\textsuperscript{18}O lines and the HAWC$+$ polarization measurements, both data sets were quantitatively compared. Unlike the almost constant magnetic field orientation, the velocity field experiences a strong local reorientation when crossing the filament crest. 
This suggests that the velocity field of the filament crest is more complex than the magnetic field, although we do see a slight ($\sim$ 3$^{o}$) local reorientation in the magnetic field from east to west of the filament.
\end{itemize}

Our work indicates that the Musca cloud features a 3D curved filament embedded in an ambient cloud with an organized magnetic field, however more observations made towards Musca are required in order to untangle both the structure of Musca and the dynamics guiding its evolution.

\begin{acknowledgments}
We thank A. Hacar for providing the APEX C$^{18}$O(2-1) pointing observations and detailed comments on a draft of the paper. We thank J. Soler for discussions on the HAWC$+$ data set. We further thank N. Schneider for useful comments on an early version of the manuscript and we also thank M. McAdam and N. Rangwala for helping to obtain the funding from the SOFIA project for the NASA internship of A.K. This work is based on observations made with the NASA/DLR Stratospheric Observatory for Infrared Astronomy (SOFIA). SOFIA is jointly operated by the Universities Space Research Association (USRA) under NASA contract NNA17BF53C, and the Deutsches SOFIA Institut (DSI) under DLR contract 50 OK 0901 to the University of Stuttgart. We thank the USRA and NASA staff of the Armstrong Flight Research Center in Palmdale and of the Ames Research Center in Mountain View, and the Deutsches SOFIA Institut for their work on the observatory. Based on observations obtained with Planck (http://www.esa.int/Planck), an ESA science mission with instruments and contributions directly funded by ESA Member States, NASA, and Canada. This publication is based on data acquired with the Atacama Pathfinder Experiment (APEX) under programme IDs 087.C-0583(A). This research has made use of data from the Herschel Gould Belt survey project (http://gouldbelt-herschel.cea.fr). The HGBS is a Herschel Key Project jointly carried out by SPIRE Specialist Astronomy Group 3 (SAG3), scientists of several institutes in the PACS Consortium (CEA Saclay, INAF-IAPS Rome and INAF-Arcetri, KU Leuven, MPIA Heidelberg), and scientists of the Herschel Science Center (HSC). L.B. was supported by a USRA postdoctoral fellowship, funded through the NASA SOFIA contract NNA17BF53C.
\end{acknowledgments}

\facilities{SOFIA (HAWC+), APEX, ALMA, Planck, Herschel}

\software{astropy \citep{Astropy2013,Astropy2018}}

\bibliography{sample631}
\bibliographystyle{aasjournal}

\end{document}